\documentclass{article}
\usepackage{hiph-preprint}
\usepackage{graphicx}
\usepackage{amssymb,amsmath,citesort}
\volnumber{22} \issuenumber{1} \edyear{2005}                             
\frompage{000} \topage{000}                                              
\recrevdate{30 October 2005}                                             

\newcommand{\eqeqref}[1]{Eq.~\eqref{#1}}

\newcommand{\refref}[1]{Ref.~\cite{#1}}

\newcommand{\figref}[1]{Fig.~\ref{#1}}

\newcommand{\beq}{\begin{equation}}
\newcommand{\eeq}{\end{equation}}
\newcommand{\bea}{\begin{eqnarray}}
\newcommand{\beas}{\begin{eqnarray*}}
\newcommand{\beau}[1]{\begin{equation} \label{#1} \begin{array}{rcl}}
\newcommand{\eea}{\end{eqnarray}}
\newcommand{\eeas}{\end{eqnarray*}}
\newcommand{\eeau}{\end{array} \end{equation}}
\newcommand{\bay}{\begin{array}}
\newcommand{\eay}{\end{array}}
\newcommand{\bals}{\begin{align*}}
\newcommand{\eals}{\end{align*}}

\newcommand{\vev}[1]{\langle #1 \rangle}

\newcommand{\PP}{{\mathcal P}}

\def\rightmark{Energy loss vs. hadron absorption}
\def\leftmark{A. Accardi}

\title{
Can we distinguish energy loss from hadron absoprtion?$^\dagger$
} 
\authors{ 
{Alberto Accardi$^1$%
\index{Accardi, A.} 
}\\[2.812mm]
{\normalsize
\hspace*{-8pt}$^1$ Department of Physics and Astronomy, Iowa State
University, \\
Ames, Iowa 50011-3160, U.S.A.\\ 
\hspace*{-8pt}$^\dagger$ Write-up of poster presented at ``Quark
Matter 2005'', Budapest, August 2005
}}
 
\abstract{
Knowing whether a hadron is formed inside
or outside the nuclear medium is very important for
correctly interpreting jet quenching in heavy-ion collisions. 
The cleanest
experimental environment to study the space-time
evolution of hadronization is semi-inclusive DIS on
nuclear targets.
2 frameworks are presently competing to explain the
observed attenuation of hadron production: quark energy
loss, with hadron formation outside the nucleus \cite{Wang,Arleo},
and nuclear absorption with hadronization starting inside the nucleus
\cite{AGMP05,Falteretal04,Kopeliovich}. 
I demonstrate that the observed approximate $A^{2/3}$ scaling of
experimental data cannot conclusively establish the correctness of 
either energy loss or absorption.
}

 
\makeindex

\begin{document}
 
\maketitle


\noindent

In Deep Inelastic Scattering on nuclear targets (nDIS) one observes a
suppression of hadron production \cite{HERMES}
analogous to hadron quenching in heavy-ion collisions. However, nDIS
offers a much cleaner experimental environment to study quark
fragmentation: the nuclear
medium is well known and the multiplicity in the final state is low.
Knowing how the struck quark propagates in cold nuclear matter
-- most importantly, whether its color is neutralized inside or outside
it -- is a necessary prerequisite for correctly using hadron quenching
data to extract the unknown properties of the hot and dense medium
produced at RHIC.  
Experimental data on hadron production in nDIS are usually presented
in terms of the ``attenuation ratio'' 
$
  R_M^h(z) = \frac{1}{N_A^{DIS}}\frac{dN_A^h(z)}{dz} / 
    \frac{1}{N_D^{DIS}}\frac{dN_D^h(z)}{dz} ,\
    \label{MultiplicityRatio}	   
$
i.e., the single hadron multiplicity on a target of mass number $A$ 
normalized to the multiplicity on a deuterium target. 
I will only analyze the dependence of $R_M^h$ 
on the hadron's fractional energy $z=E^h/\nu$, but data
also exist binned in the virtual photon energy $\nu$ or
its virtuality $Q^2$. 
\\

\noindent 
In the hadron absorption model of \refref{AGMP05} hadronization is
assumed to happen in 2 stages: (i) the struck quark neutralizes its
color and forms a so-called ``prehadron'' $h^*$, and (ii) the observed
hadron $h$ is formed. The average formation length of the prehadron,
$\vev{l^*}(z,\nu)$, and of the hadron, $\vev{l^h}(z,\nu)$, are computed in the
framework of the standard Lund model. In the HERMES kinematics
\cite{HERMES} the
prehadron is formed well inside the nucleus while the hadron is
produced mostly outside. After its formation, the prehadron
can interact with the surrounding nucleons with a cross section $\sigma^*(\nu)
= 2/3 \, \sigma^h(\nu)$ proportional to the experimental hadron-nucleon cross
section $\sigma^h$. The proportionality factor is fitted to 
$\pi^+$ production data on a Kr target at $E_\text{beam} = 27$ GeV$^2$
\cite{HERMES}.
The probability $S^A_{f,h}(z,\nu)$ that neither
the prehadron nor the hadron interact can be computed using
transport differential equations \cite{AGMP05}:
\begin{align} 
  S_{f,h}^A(z,\nu) = & \int d^2b\,dy\,\rho_A(b,y)  \nonumber \\
  & \times 
    \int\limits_y^{\infty}dx'\int\limits_{y}^{x'}dx\, 
    \frac{e^{-\frac{x-y}{\left< l^* \right>}}}
    {\left< l^*\right>}e^{-\sigma_*\int\limits_x^{x'}dsA\rho_A(b,s)}\,
    \frac{e^{-\frac{x'-x}{\left< \Delta l \right>}}}
    {\left< \Delta l\right>} 
    e^{-\sigma_h\int\limits_{x'}^{\infty}dsA\rho_A(b,s)}
\end{align}
where $\Delta l = l^h-l^*$, and $\rho_A$ is the nuclear density.
One can recognize exponential probability
distributions for prehadron and hadron formation.
Note also the 
integration over the interaction points $(b,y)$ of the virtual photon
$\gamma^*$ with the quark. The hadron multiplicity is 
computed, at leading order in pQCD, as follows:
\begin{align} 
  \frac{1}{N_A^{DIS}}\frac{dN_A^h(z)}{dz} =\; &  
       \frac{1}{\sigma^{lA}} \hspace{-0.2cm}
       \int\limits_{\mbox{\footnotesize exp. cuts}}
       \hspace{-0.4cm}
       dx\,d\nu\,
   \sum_f e_f^2 q_f(x,Q^2) 
      \frac{d\sigma^{lq}}{dx d\nu} S_{f,h}^A(z,\nu) D_f^h(z,Q^2) \ .
 \label{eq:DISxsec}
\end{align}
Here $\sigma^{lq}$ and $\sigma^{lA}$ are the lepton-quark and
lepton-nucleus cross sections. $q_f$ is the
$f$-quark distribution function, and $D_f^h$ its fragmentation
function.
The model neglects hadron elastic scatterings, 
diffractive hadron production, and resonance production and
decay. Therefore it is applicable at $0.4 \lesssim z \lesssim
0.9$.  

Energy loss models \cite{Wang,Arleo} assume that the struck quark hadronizes
well outside the medium, and that hadron attenuation is due to medium-induced
gluon radiation off the quark. The 
quark's energy is reduced from $E_q = \nu$ to
$E_q=\nu-\epsilon$, where $\epsilon$ is the total energy of the
radiated gluons, which translates into a modified fragmentation function:
\begin{align*} 
  \tilde D_f^h (z,Q^2;L) = & \int\limits_0^{(1-z)}\!\!\!\! d\Delta z\; 
    \PP(\Delta z\,;\hat q,L) \frac{1}{1-\Delta z}
    D_f^h(\frac{z}{1-\Delta z},Q^2) 
  + p_0(\hat q, L) \, D_f^h(z,Q^2) \ .
\end{align*}
The ``quenching weight'' $\PP(\Delta z)$ 
is the probability distribution of a fractional energy loss $\Delta
z=\varepsilon/\nu$, and $p_0$ is the probability of zero energy loss
\cite{SW03}. The quenching weight depends on the quark's in-medium
path length $L_A(b,y)$ and on the transport coefficient $\hat 
q(b,y)$, defined to take into account non-uniform nuclear
density \cite{Accardiextra}. $\hat q(0,0) = 0.5$ GeV$^2$/fm,
is fitted to $\pi^+$
production on a Kr target analogously to the absorption model.
Hadron multiplicity is then computed as in \eqeqref{eq:DISxsec},
using the modified $\tilde D$ and $S^A=1$. Finally, an
integration over $(b,y)$ is performed.

\begin{figure}[tb]

  \vspace*{0cm}
  \centerline{
  \includegraphics[height=5.4cm]{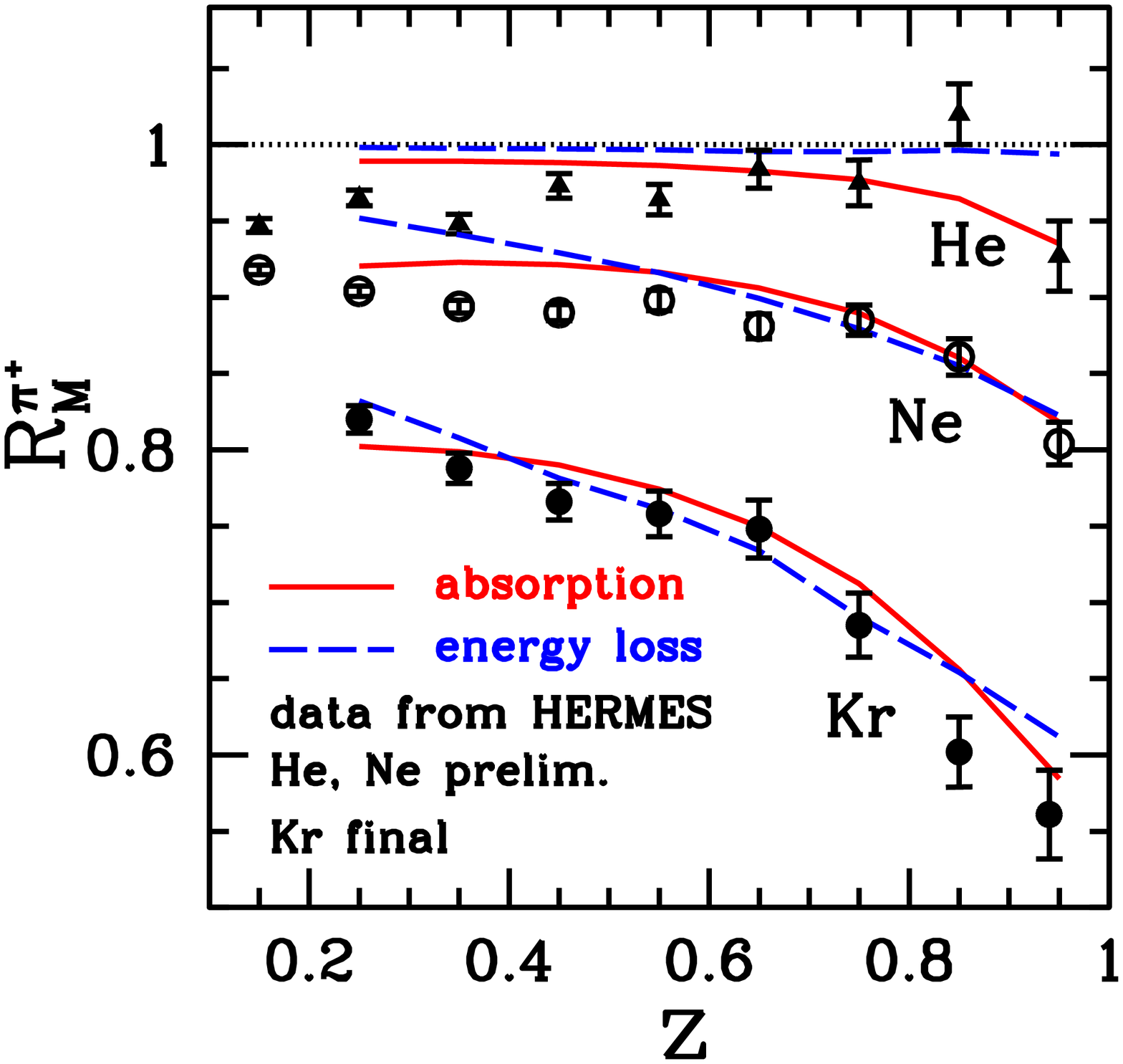}\ \
  \includegraphics[height=5.4cm]{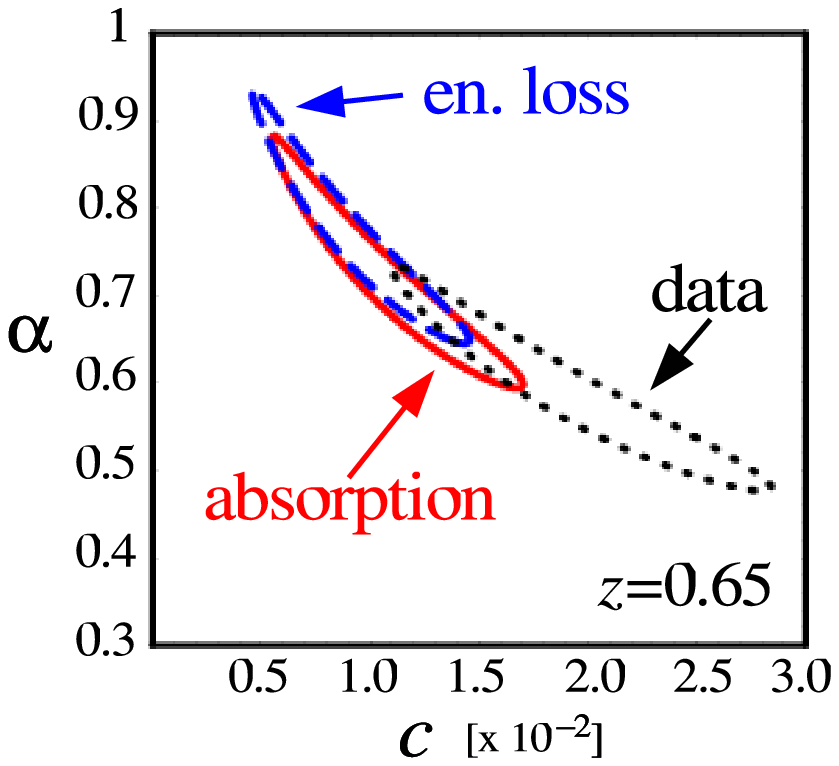}
  }
  \vspace*{-.4cm}
 \caption[]{
  Left: $\pi^+$ multiplicity ratio in the absorption and energy loss
  models compared to HERMES data at a beam energy of 27 GeV$^2$
  \cite{HERMES}. 
  Right: results of the $R_M=cA^\alpha$ fit for \{He,N,Ne,Kr\} at
  z=0.65 (solid: absorption; dashed: energy loss; dotted: data \cite{HERMES}). 
  More $z$-bins and case studies can be found in \cite{Accardiextra}.}
 \label{fig:RMplus}
\end{figure}

In \figref{fig:RMplus}, the 2 models are compared to HERMES data. 
Both describe well the data, and look similar despite the
different physics processes. Quite surprisingly, this similarity holds
up to very heavy targets like Pb.  
\\


\noindent 
A na\"\i ve argument, often repeated in discussions and seminars on
heavy-ion collisions, is that in first approximation  $1-R^h_M \propto
A^{2/3}$ in energy loss models because the average energy loss
$\vev{\epsilon} \propto \vev{L_A^2} \propto A^{2/3}$.
On the other hand,
in absorption models the survival probability is proportional to the
amount of traversed matter, so that $1-R^h_M \propto \vev{L_A}
\propto A^{1/3}$.
Therefore, it is concluded, a simple analysis of the $A$-dependence of
$R_M^h$ (or of $R_{AA}^h$ in heavy-ion collisions) will clearly signal
which one of the 2 models is correct. 

The above argument is wrong! Where the argument actually fails is
for absorption models \cite{enlossargument}. If the 
prehadron were produced always at the $\gamma^*$-quark interaction
point (i.e., $\vev{l^*}=0$) then $R_M = c \,A^{1/3}$ at all
orders in $A^{1/3}$. However, if we allow for a nonzero $\vev{l^*}$, 
its dimensions must be neutralized by the nuclear radius $R_A$,
introducing extra powers of $A^{1/3}$. Quite generally, if the
probability distribution for the prehadron formation 
length is finite at zero formation length, then $R_M^h \propto A^{2/3} +
O(A)$, the same power found in energy loss models  \cite{AGMP05}. 
This is the case for the presented model, as well as
for most other absorption models.

Then, we can hope to distinguish energy loss from hadron absorption
by studying the {\em breaking} of the $A^{2/3}$ law. To this
purpose, it was proposed in \cite{AGMP05} to select a set of targets
$\{A_1,A_2,\ldots,A_n\}$, fix the $z$ bin, 
and perform a fit of the form $1-R^h_M(z) =
c(z) A^{\alpha(z)}$. Both $c$ and $\alpha$ must be considered fit
parameters for 2 reasons: first, both contain information on the
dynamics of the hadronization process \cite{AGMP05,Accardiextra}; 
second, one can always redefine $c$ in order to
absorb a part of $\alpha$ biasing the result, so that it is more
correct to ask the fit itself what are the correct values of the 2
parameters. The results of the fit are presented in terms of $2\sigma$
confidence contours in the $(c,\alpha)$ plane, see
\figref{fig:RMplus}. 
This analysis is
powerful: it is sensitive to model parameters like $\hat q$
\cite{Accardiextra}, and to different physical mechanisms: e.g., 
partial quark deconfinement in nuclei \cite{AGMP05}.
However, when we apply the fit to the 2 models described in the
previous section, we have a surprise: energy loss and absorption are
indistinguishable. The same holds true
for all $z$ bins. Increasing the number of targets and the span in
atomic number does not help in separating the 2 models, either, but clearly
shows a non negligible breaking of the $A^{2/3}$ law at $A \gtrsim 80$
\cite{Accardiextra}.  
\\


\noindent 
In summary, single hadron suppression obeys a $A^{2/3}$ law (broken at
$A \gtrsim 80$) in both energy loss and absorption models. Thus, the
observed approximate $A^{2/3}$ scaling of experimental 
data for light nuclei cannot be used as a proof of the energy loss
mechanism, as is often done. 
Even the more refined analysis 
in terms of $(c,\alpha)$ fits cannot clearly distinguish
the 2 classes of models, though it will 
help in checking the details of the ``correct'' model after this is
established by other means.
To answer the very important question of whether or not 
the struck quark starts hadronizing inside the nucleus, we need to
consider more exclusive observable, like the $z$-dependence of  hadron's
$p_T$-broadening and Cronin effect \cite{Kopeliovich} or dihadron
correlations \cite{HERMESdihadron}.

\section*{Acknowledgments}
A special thank to D.Gr\"unewald for his help in the initial phase
of this investigation. This work is partially funded by the US
Department of Energy grant DE-FG02-87ER40371.

\vfill\eject

\begin{thebibliography}{99}
\bibitem{enlossargument}
  In energy loss models, a leading order  $A^{2/3}$ term can be
  obtained only by neglecting finite-medium corrections and
  energy conservation, $\Delta z < (1-z)$. Nonetheless, for light
  nuclei up to Kr the $2/3$ power is approximately correct.
\end{thebibliography}
\end{document}